\documentclass[10pt,twocolumn]{article}

\usepackage{graphicx}
\usepackage{amsfonts}
\usepackage{amsmath}
\usepackage{amssymb}
\usepackage{mathrsfs} 
\usepackage{lipsum}
\usepackage{color}
\usepackage{authblk}
\usepackage[margin=2cm]{geometry}
\usepackage{mathtools, cuted}
\usepackage{bm}
\usepackage{amsmath}
\usepackage[utf8]{inputenc}
\usepackage[squaren,Gray,cdot]{SIunits}
\usepackage{caption}
\usepackage{subcaption}
\usepackage{mathtools, cuted}
\usepackage{array,amsmath}
\usepackage{textcomp}

\begin{document}
\title{Stability of Big Surface Bubbles: Impact of Evaporation and Bubbles Size}
\author{Jonas Miguet$^1$, Marina Pasquet$^1$, Florence Rouyer $^2$, Yuan Fang $^3$, Emmanuelle Rio$^1$\\
\small{$^1$ Univ. Paris Sud, Laboratoire de Physique des Solides, CNRS UMR 8502\\
$^2$ Laboratoire Navier, Université Paris-Est, 77454 Marne-la-Vallée, France\\
$^3$ PepsiCo Global R\&D, Valhalla, New York 10595, United States}
}
\maketitle

\begin{abstract}
Surface bubbles have attracted much interest in the past decades. In this article, we propose to explore the lifetime and thinning dynamics of centimetric surface bubbles. We study the impact of the bubbles size  as well as that of the atmospheric humidity through a careful control and systematic variation of the relative humidity in the measuring chamber. We first adress the question of the drainage under saturated water vapor conditions and show that a model including both capillary and gravity driven drainage provides the best prediction for this process. Additionally, unprecedented statistics on the bubbles lifetimes confirm experimentally that this parameter is set by evaporation to leading order. We make use of a model based on the overall thinning dynamics of the thin film and assume a rupture thickness of the order 10-100 nm to obtain a good representation of these data.  For experiments conducted far from saturation, the convective evaporation of the bath is shown to dominate the overall mass loss in the cap film due to evaporation.
\end{abstract}

\section{Introduction}

Due to their wide range applicability, surface bubbles have attracted considerable attention is the past decades. The general reasons for this is their ubiquity and the enhanced transfer of materials from the liquid reservoir to the overlying atmosphere through the production of aerosols during the burst. In societal applications, studies may be found in different contexts: domestic\cite{Johnson2013}, recreational\cite{Embil1997}, industrial\cite{Beerkens2006}. Because surface active materials may adsorb at their surface during the ascent \cite{Bond1928}, the produced aerosols feature excess concentration of such materials, which has consequences in the release of flavours from fizzy drinks \cite{Liger-Belair2009}. In a geophysical context, the aerosols produced by bursting bubbles at the surface of the oceans constitute a primary natural source \cite{IPCC_AR5_CH7} that influences clouds formation and their radiative properties \cite{murphy1998influence}. They can also favour the transport of pathogens \cite{baylor1977virus}, which, in turn may alter the bubble stability \cite{Poulain2018a}.

The number of produced aerosols, their size and ejection velocity depend on the film thickness, which in turn is linked both to the bubble lifetime and to the thinning dynamics of the thin liquid film, which are the objects of this study. Typical scenario for the life of a surface bubble, that we detail further, is: emergence at the air/liquid interface, adoption of an equilibrium shape at the surface, thinning of the cap film, nucleation of a hole and subsequent bursting. 

When a bubble emerges at a liquid/gas interface, {it undergoes several bounces that typically last microseconds to hundreeds of microseconds in the case of milimetric bubbles, as was shown by Zawala \textit{et al.} \cite{zawala2011influence}. A first regime of fast thinning of the bubble cap film proceeds until it adopts an equilibrium shape that is determined by the balance of buoyancy, that pulls the gas phase in the volume towards the atmosphere and the surface tension-induced force along the circular meniscus that binds it to the bath \cite{Teixeira2015}. Therefore, a relevant dimensionless quantity to address the question of the shape of the bubble is the Bond number :
\begin{equation}
	Bo = \frac{\rho_{\text{liq}} g{R}^2}{\gamma},
\end{equation}
where $\rho_{\text{liq}}$ [kg.m$^{-3}$] and $\gamma$ [N.m$^{-1}$] are respectively the density and the surface tension of the liquid (provided that the density of the gas may be neglected), $g$ the acceleration due to gravity [m.s$^{-2}$], and $R$ [m] the radius of the spherical cap of the bubble, above the meniscus. After it has reached its final shape, the thin film of the spherical cap undergoes thinning. Different scenarii have been proposed in the litterature to describe this step. 
For pure viscous liquids, Debrégeas et al. \cite{Debregeas1998} proposed a gravity driven flow with zero interfacial stress. The viscous film thickness on top of the bubble decays exponentially with a characteristic time related to the shape of the bubble \cite{pigeonneau2011low, kovcarkova2013film}.
For liquids with surface active species, a metastable equilibrium is reached, where the surface tension gradient necessary to hold the weight of the film is established. A slower thinning mechanism develops, driven by capillary pressure and/or gravity, but also by evaporation \cite{Poulain2018a}. 
A continuous transition from zero stress to zero velocity boundary condition (depending on the relative strength of interfacial to bulk stresses) was proposed by Bhamla \textit{et al.} \cite{Bhamla2014}. Champougny \textit{et al.} \cite{Champougny2016} measured different thinning rates for different surfactant concentrations and proposed a model based on an intermediary boundary condition \textit{ie} a slip length. Lhuissier \textit{et al.} proposed a drainage model that will be discussed further, based on the coupling between a capillary driven flow localised at the foot of the bubble and the periodic emission of so called marginal regeneration plumes, as first reported by Mysels \textit{et al.} \cite{Mysels1959}

Another potential contributor to thinning is evaporation, about which literature is much more scarce with only one recent paper to address its impact on the stability of surface bubbles \cite{Poulain2018a}. However, the few systematic studies of the stability of thin films under controlled partial pressure of water in the gas phase, relative to the saturation pressure (\textit{ie} relative humidity) report a strong impact of this parameter \cite{Li2010,Champougny2018,Pagureva2016}. Bubble artists also invariably report decreasing stability of their films with decreasing humidity. On the other hand, when the evaporation rate gets high enough, it can lead to Marangoni flows that can stabilise the bubble \cite{Poulain2018}.

As the film thins, it becomes more and more prompt to nucleate a hole that expands quickly (typically milliseconds\cite{spiel1998births}) and irremediably leads to the destruction of the bubble. The process that initiates the rupture is not fully understood \cite{Rio2014} but the stochastic nature of this event is well established by now \cite{haffner2018can, Forel2019a}. The total lifetime of the bubbles being the sum of both contribution (thinning and initiation of a hole), a sufficient characterisation of a given system can only be achieved with repeated measurements.

In this paper, we provide the first study of the stability of surface bubbles with a systematic variation of their size (here centimetric bubbles) in a controlled humidity environment. We benefit from an automated generation and measurement of the bubble lifetime to obtain good statistics  (thousands of bubbles). We show that the characteristic time relevant to predict the bubble lifetime is given by the comparison between the drainage velocity and the evaporation rate.

\section{Experimental methods}
\label{sec:ExpMeth}
The physicochemical system used for this study consists of ultrapure water (resistivity=18.2~M.$\Omega$.m) to which 0.5~cmc (\textit{ie} 0.62~g.L$^{-1}$) \cite{Bergeron1997} of Tetradecyl Trimethyl Ammonium Bromide, thereafter referred to as TTAB, is added. The latter is purchased from Sigma Aldrich and is further purified through recrystalization \cite{Stubenrauch2005}. It was indeed found that, at this low concentration, the interface is progressively polluted (likely by the traces of tetradecanol left after the original synthesis), which drastically lowers the equilibrium surface tension and therefore alters the reproductibility of the experiments. 

\begin{figure}[!ht]
  \centering
    \includegraphics[width=.8\linewidth]{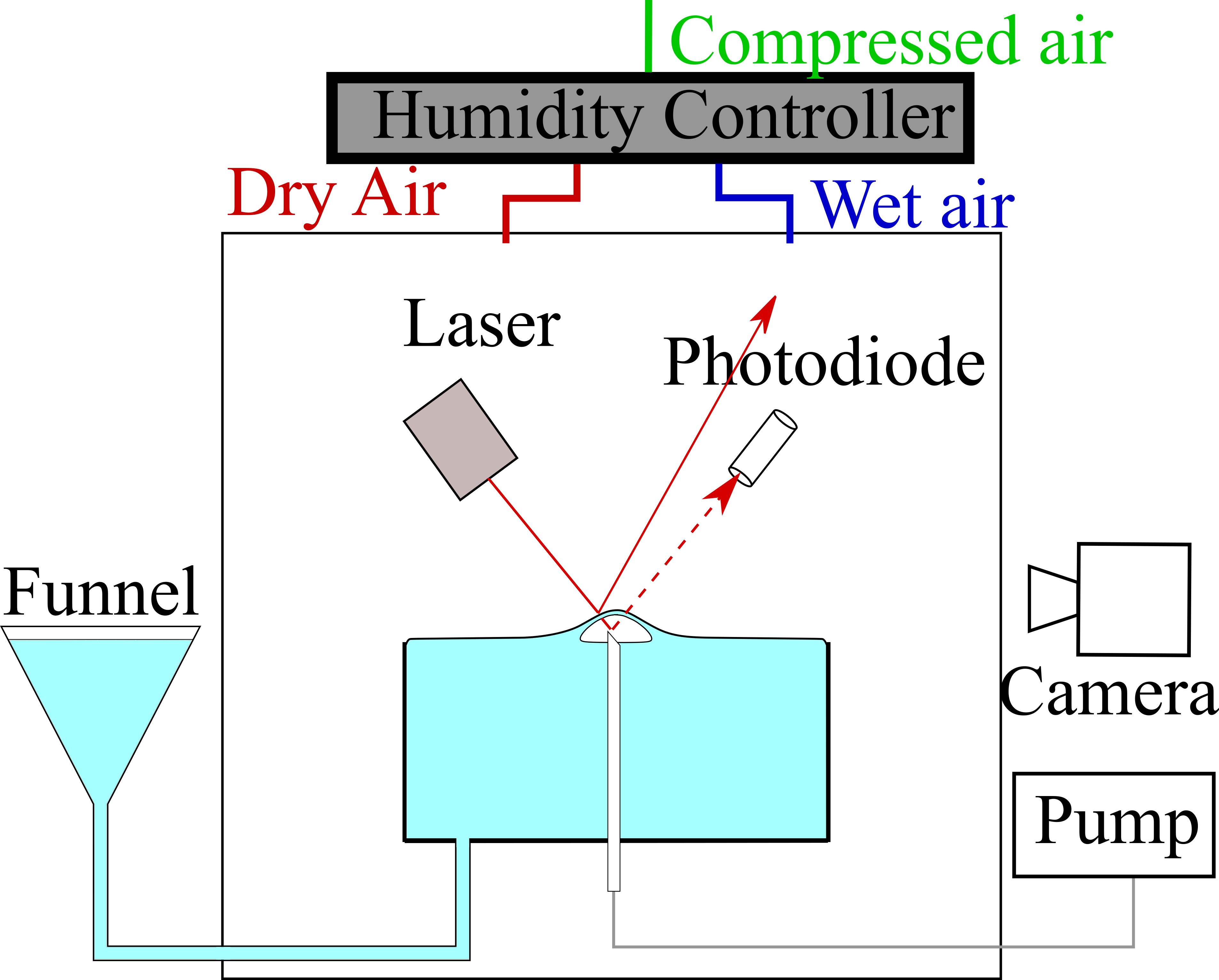}
  \caption{A container with the solution of interest is set in a humidity controlled chamber. The container is filled from the outside of the chamber, and the level of liquid is precisely controlled with a funnel, trough hydrostatic adjustment to ensure a proper laser alignment in absence of a bubble (dotted line). Air is injected by a pump to create the bubbles, the presence of which is assessed by the subsequent divergence of the laser beam with respect to the photodiode (solid line). Images are recorded from the side during the experiments.}
  \label{fig:Fig1}
\end{figure}

\begin{figure*}[!ht]
  \centering
    \includegraphics[width=1\textwidth]{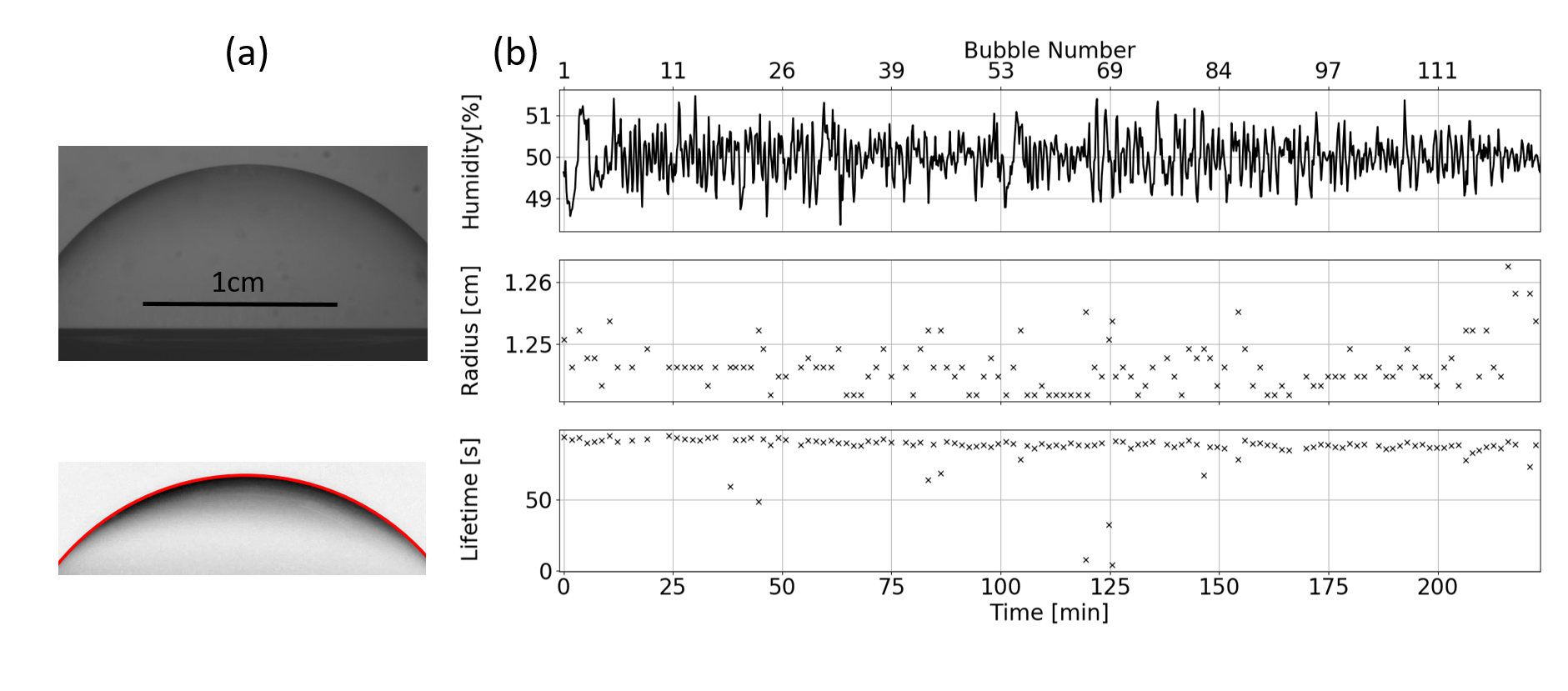}
   \caption{(a) Top: raw image of a bubble which spherical cap radius is 1.25 cm. Bottom: the same image after processing with scikit-image library in Python. The red arc of circle represents the optimal fit to the contour of the bubble and provides a measure of the radius of curvature of the cap.(b) Raw experimental data for each bubble (top axis) along time (bottom axis). Top: humidity measured in the chamber. Middle: radius of the bubbles. Bottom: lifetime of the bubbles. }
  \label{fig:Fig2}
\end{figure*}

The experimental set-up for the measurement of the bubbles lifetime is represented in figure \ref{fig:Fig1}. A cylindrical container featuring a teflon ring in its upper end is used to facilitate the emergence of a meniscus above the level of the container and images of the bubbles (see figure \ref{fig:Fig2}.(a)) are taken from the side with a monochrome camera (Marlin). A red laser beam (wave length and light power are respectively equal to of 650 nm and 5 mW) is directed to the center of the container, where the bubbles are created. In the absence of a bubble, the reflection of the beam on the interface is focused with a convex lens on a photodiode that emits a subsequent electric signal. When a bubble appears, the beam diverges and the photodiode is turned off. Making use of a Python routine, this system allows for the repeated and automated generation of bubbles together with a measurement of bubbles lifetime. The whole set-up is set in a 75x45x45 cm$^{3}$ plexiglas chamber. A humidity sensor (SHT25) is placed a few centimeters away from the bubble in the chamber, on the same horizontal plane and coupled to a flow regulator. A PID controller determines whether the output flow of air passes directly to the chamber (dry air), or first through the bottom of a water bottle (humid air), which allows for the control of the humidity RH in the measuring chamber. The room is kept at a constant temperature of 22$^{\circ}$C. In order to precisely control the level of the meniscus, the container is plugged to a funnel that is placed out of the box and which vertical position sets that of the meniscus making use of hydrostatic equilibrium. The images are taken at 4 frames per second and processed \textit{a posteriori} making use of the scikit-image library in Python. This provides an independent measurement of the bubbles lifetime, necessary to eliminate some artefacts that can arise when daughter bubbles \cite{Bird2010} or a loss of height caused by evaporation prevent the alignment of the laser with the photodiode. The size of every bubbles is also measured using this routine (figure \ref{fig:Fig2}.(a)). For the injection of the bubbles, a PTFE tube of inner diameter 325~$\mu$m is guided trough a capillary tube to which it is hermetically glued, from the bottom of the container. The lower end of the PTFE tube is plugged to a solenoid valve. It allows a flow of air triggered by a flow-controlled aquarium pump to blow bubbles when required, for a controlled amount of time, which sets the size of the bubbles in a reproducible fashion.

Figure \ref{fig:Fig2}b represents the time evolution of the different parameters in the course of an experiment. The bottom axis indicates the time ellapsed since the onset of the set up, while the top axis displays the corresponding bubble number. The top chart indicates that the humidity is properly controlled within 1.5\%. This is to compare with the accuracy of the humidity sensor, which is around 1.8 \% below  90 \%, around 2 \% between 90 and 95 \% and around 2.5\% above 95 \%. The accuracy of the humidity control is thus limited by the sensor and decreases with the humidity value. The experiments conducted at high humidity (\textit{ie} close to saturation) thus  could not be as precisely controlled in terms of absolute humidity because the precision of the sensors drops and condensation could damage the electronic connections. However, humid air was continuously injected in the chambers, and condensation could be observed during these experiments, while the few readings we took guaranteed values above 95\%.

The middle chart in Figure \ref{fig:Fig2}.(b) shows that, for a given injection time the size of the bubbles is reproducible ($\pm$ 40~$\mu$m). The bottom one shows that, for a given humidity in the chamber and a given bubble size, the bubbles lifetime does not vary significantly, which is necessary to assess an overall reliable significance of the measured lifetimes \cite{Poulain2018}.

To provide more insight into these systems, the thickness evolution of the film at the apex of the bubbles was measured using a reflectometric technique, described in more details in \cite{Champougny2016,Champougny2018}. The upper part of the film is illuminated with a white light source. A spectrometer placed vertically emits and collects the light, which ensures that the reflected light comes from a zone close to the apex of the bubble. The reflected light is recorded and analysed by the spectrometer, providing spectra of the light intensity as a function of the wavelength. The interferences between the light reflected on each side of the film lead to oscillations of the spectra, which are analysed making use of a semi automated procedure available in the form of a Python code called oospectro. The code makes use of the fact that the spectrum  is a $\frac{2\pi n h}{\lambda}$ periodic function.The maxima and minima of the recorded spectra are detected and the $h$ that fits the better the data is extracted. We then verify every spectra manually since some of them exhibit artefacts, possibly because of small vibrations or dusts.
These experiments were conducted in a closed chamber which humidity was controlled with another device described in \cite{boulogne2019}. The reproducibility of the measurements was ensured by measuring the same systems several times and making sure that the dynamics are the same. The error bars in the inset of Figure \ref{fig:Fig4} come from the standard deviation obtained from 4 different measurements performed in the same conditions (RH $\sim$ 100\%, R=11.3 mm, n=-0.50$\pm$0.05).

The measure of the film thickness at the apex of the bubble cannot necessarily be taken as the mean thickness in the whole film. Indeed, film thickness inhomogeneities are reported in various systems, from horizontal circular films where dimples can appear \cite{joye1992dimple,joye1994asymmetric,manev1997effect} to other processes like marginal regeneration, observed in planar films\cite{Mysels1959} as well as in bubbles\cite{Lhuissier2012}. 
Nevertheless, we make the hypothesis that the value measured at the apex is representative to analyse the cap film drainage. This is supported by the constant receding velocity of the film after the nucleation of a hole measured on metric bubbles in the work of Cohen \textit{et al.} \cite{Cohen2017}. Similar measurements have been performed on millimetric bubbles\cite{Poulain2018,Poulain2018a,Lhuissier2012}.

\section{Results and discussion}
\subsection{Results on the bubbles lifetime}

\begin{figure}[!ht]
  \centering
    \includegraphics[width=1\linewidth]{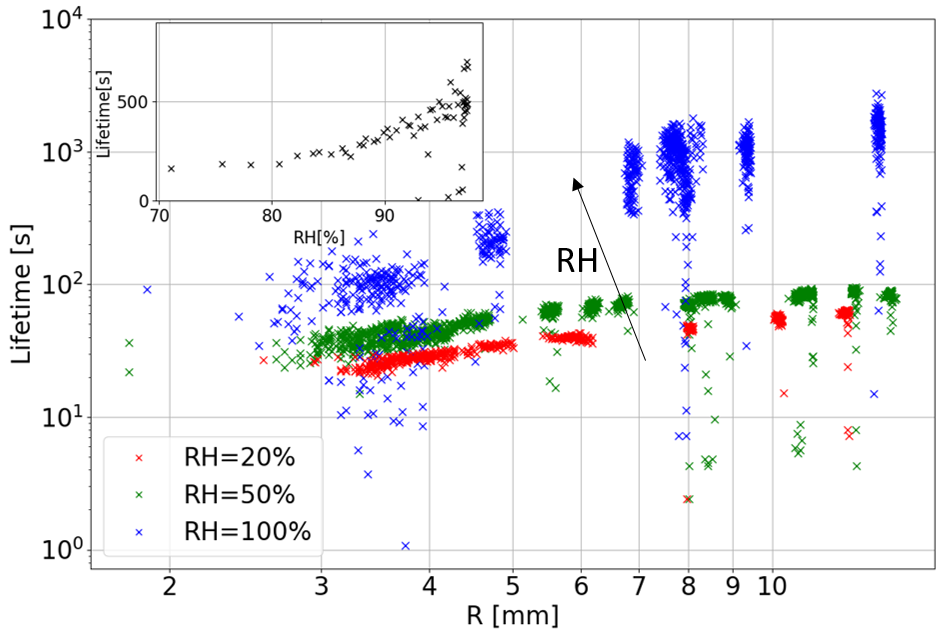}
  \caption{Raw data of bubbles lifetime as a function of their spherical cap radius of curvature for three different humidity values. The inset represents the lifetime of bubbles of radius R$\approx$1.3 cm as a function of the relative humidity in the chamber.} 
    \label{fig:Fig3}
\end{figure}

All the collected data on the bubbles lifetimes are represented in figure \ref{fig:Fig3} as a function of their size, for different humidity values. The stability of the bubbles increases significantly with their size for a given humidity. On the other hand, environmental humidity is demonstrated to play a crucial role in the stability of the bubbles. We indeed measured lifetime differences up to one order of magnitude for similar bubbles submitted to different relative humidity values. This effect is not linear since the stability increases approximately by a factor of 10 between 50\% and saturated values, while it increases only by a factor of two between 20 \% and 50 \%. A complementary experiment to further document this behaviour is performed by fixing the bubbles size at 1.3 cm and letting the humidity value raise from 65 \% to 99 \% (see inset). 

In Figure \ref{fig:Fig3}, the dispersion of the data points seems to increase with the relative humidity. This may be due to the accuracy of the humidity control, which decreases with RH, as mentioned in section \ref{sec:ExpMeth}. However, some random early burst events are recorded for all sizes and humidities that lead to up to two orders of magnitude differences for the lifetime of otherwise equivalent systems. This point is generally reported in all systematic studies of thin films stability, as shown for instance by Poulain \textit{et al} \cite{Poulain2018a} where spontaneous bursts are reported on the same system for thicknesses up to almost 10 $\mu$m, and yet to be understood. However clear trends can be observed in the maximal lifetime of the bubbles as a function of their size and environmental humidity.

\subsection{Bubbles drainage}
\label{subsec:drainage}
\begin{figure}[!ht]
  \centering
    \includegraphics[width=1\linewidth]{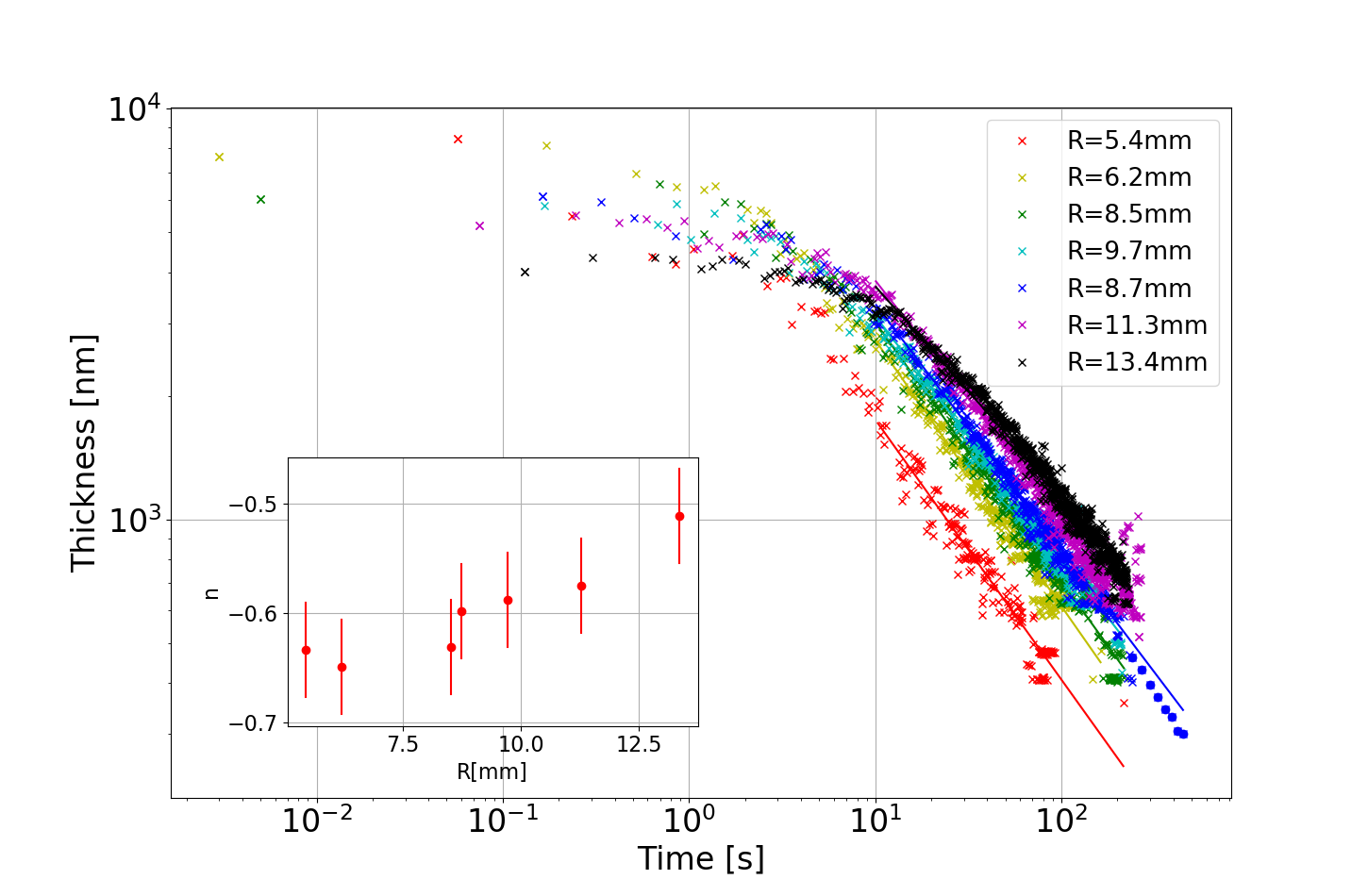}
  \caption{Thickness of bubbles film at the apex as a function of time, for various bubbles size, close to water vapor saturation. Full lines are the optimal fits using a power law with the exponent as a free parameter.The plain circles are obtained using the software NanoCalc to process the spectra. The inset shows the dependency of this exponent with the bubbles size. To estimate the error bars, we have been fitted the dynamics on 4 different bubbles in the same conditions and calculated the standard deviation.}
    \label{fig:Fig4}
    \end{figure}

In Figure \ref{fig:Fig4}, we show the evolution of the film thickness at the apex of the bubbles versus time for an environmental humidity close to saturation for different bubbles radii (corresponding to Bond number above 5). In this situation, in (almost complete) absence of evaporation, the two potential drivers for the thinning of the cap are capillary pressure and gravity. For pure solutions, it was shown that the transition between a capillary dominated regime (with a macroscopic driving force of $2\pi\gamma S/R$, where $S$ [m$^{2}$] is the area of the bubble cap) and a gravity dominated regime (where the driving force is $4\pi R^{3}\rho_{\text{liq}} g/3$) occurs for a Bond number of 0.25 \cite{nguyen2013film}. Our system is however qualitatively different because the presence of surfactants can induce a so-called marginal pinching in the vicinity of the bubble foot (that is, at the transition region between the meniscus and the overlaying cap). This pinch was proved by Aradian \textit{et al.} \cite{aradian2001marginal} theoretically consistent with rigid boundary conditions at the surface of the films, while Howell \textit{et al.} \cite{howell2005absence} demonstrated the impossibility for a film to pinch in the case of fully mobile interfaces. In the case of TTAB solutions, neither rigid nor mobile film thinning models would prove consistent with experimental observations \cite{Champougny2016}. However, an indirect proof of the existence of such pinching can be obtained through the observation of convective plumes, of smaller thickness, that rise from the bottom to the cap of the bubble. This phenomenon has been reported in numerous studies both in the case of vertical films \cite{Mysels1959,Nierstrasz1998,Seiwert2017} and bubbles \cite{Champougny2016,Lhuissier2012}. Lhuissier et. al\cite{Lhuissier2012} considered for the first time the influence of the pinch on the overall cap film drainage dynamics. Featuring the smallest thickness (and therefore maximum viscous dissipation), the pinched zone is assumed to be limiting in the whole drainage process. The surface rigidity is ensured by assuming an accumulation of surfactants at the foot of the bubble, under the action of drainage of the cap, which creates a subsequent Marangoni stress. The instability that gives rise to the plumes and the corresponding contribution to the thinning (by replacement of thick films portions by thin regeneration plumes), being dependent on the drainage driven accumulation of surfactants, is assumed to be of the same order of magnitude. Finally, considering that the thickness difference between the rising plumes and the mean cap thickness is of the same order than the cap thickness itself, they use scaling laws for the capillary flow in the limit of small bubbles within the pinch and the matching of the curvature of the cap with that of the pinch (see equation \ref{curvature_matching}) to obtain the following prediction for the cap film thickness \cite{Lhuissier2012}: 
\begin{equation}
	h \sim l_{c} \left(\frac{\eta l_{c}}{\gamma t} \right)^{2/3} \left( \frac{R}{l_{c}} \right)^{7/3},
\end{equation}
where $h$[m] is the mean thickness of the spherical cap film, $\eta$[kg.m$^{-1}$.s${^-1}$] the bulk viscosity of the solution and $l_{c}$ [m] the capillary length of the system defined as $\sqrt{\gamma / \rho_{\text{liq}} g}$. 

This predicts a scaling of the film thinning with time $h\sim t^{-2/3}$.
On the other hand, in the case of large bubbles \textit{ie} gravity driven flows, the thickness is expected to evolve as $t^{-1/2}$ whatever the boundary condition at the interfaces \cite{Bhamla2014}. 
The data in Figure \ref{fig:Fig4} indeed exhibit an algebraic behavior with time after a few seconds and we thus fitted the data by a power law.
The extracted exponent is reported in the inset of Figure \ref{fig:Fig4} and exhibits a transition from a $-2/3$ exponent for the smallest bubble to $-1/2$ for the largest, implying that both phenomena need to be taken into account for our experiment. 
Momentum conservation in the pinched area can be written using the Stokes equation, which scales as:

\begin{equation}
	\eta \frac{V}{\delta ^{2}} \sim \frac{\gamma}{Rl}+\rho_{\text{liq}} g,
	\label{MomentumConservation}
\end{equation}
\noindent where $V$[m.s$^{-1}$] is the typical velocity of the fluid within the thickness of the film, $\delta$[m] and $l$[m] the two characteristic lengths of the pinch that are, respectively, its thickness and its tangential extension (with respect to the local bubble surface, see the inset of Figure 5). Mass conservation writes:
\begin{equation}
    \frac{dh}{dt}+\frac{P}{S}hV \sim 0,
    \label{mass_conservation}
\end{equation}

where $P$[m] and $S$[m$^{2}$] are repectively the perimeter of the (circular) meniscus and the surface area of the cap. 

We now follow the same steps than Lhuissier \textit{et al.} \cite{Lhuissier2012} for the closure of the problem in presence of gravity in equation \ref{MomentumConservation}. Matching the curvature of the cap with that of the pinch implies:
\begin{equation}
	\frac{1}{R} \sim \frac{h-\delta}{l^2}
	\label{curvature_matching}
\end{equation}
The thickness of the pinch is of the order of $h$ and evolves in parallel with it, an affirmation confirmed by Nierstrasz \cite{Nierstrasz1998} who finds a constant ratio $\delta/h$ of 0.2 during the whole draining process of a vertical foam film. Equation \ref{curvature_matching} thus allows to express $l$ as the geometrical mean of the two other lengths of the problem, \textit{ie} $l\sim \sqrt{Rh}$, so that:
\begin{equation}
	V\sim \frac{\gamma}{\eta} \frac{h^{3/2}}{R^{3/2}}+\frac{\rho_{\text{liq}} gh^{2}}{\eta},
	\label{velocity}
\end{equation}
where the first term in the right-hand side accounts for capillary suction and the second one for gravity driven drainage. Finally, we assume a large bubble limit for the geometrical factor in equation \ref{mass_conservation}, namely: $P/S\sim 1/R$ in line with a Bond number larger than $5$ \cite{Teixeira2015}.
\begin{figure}[!ht]
  \centering
    \includegraphics[width=1\linewidth]{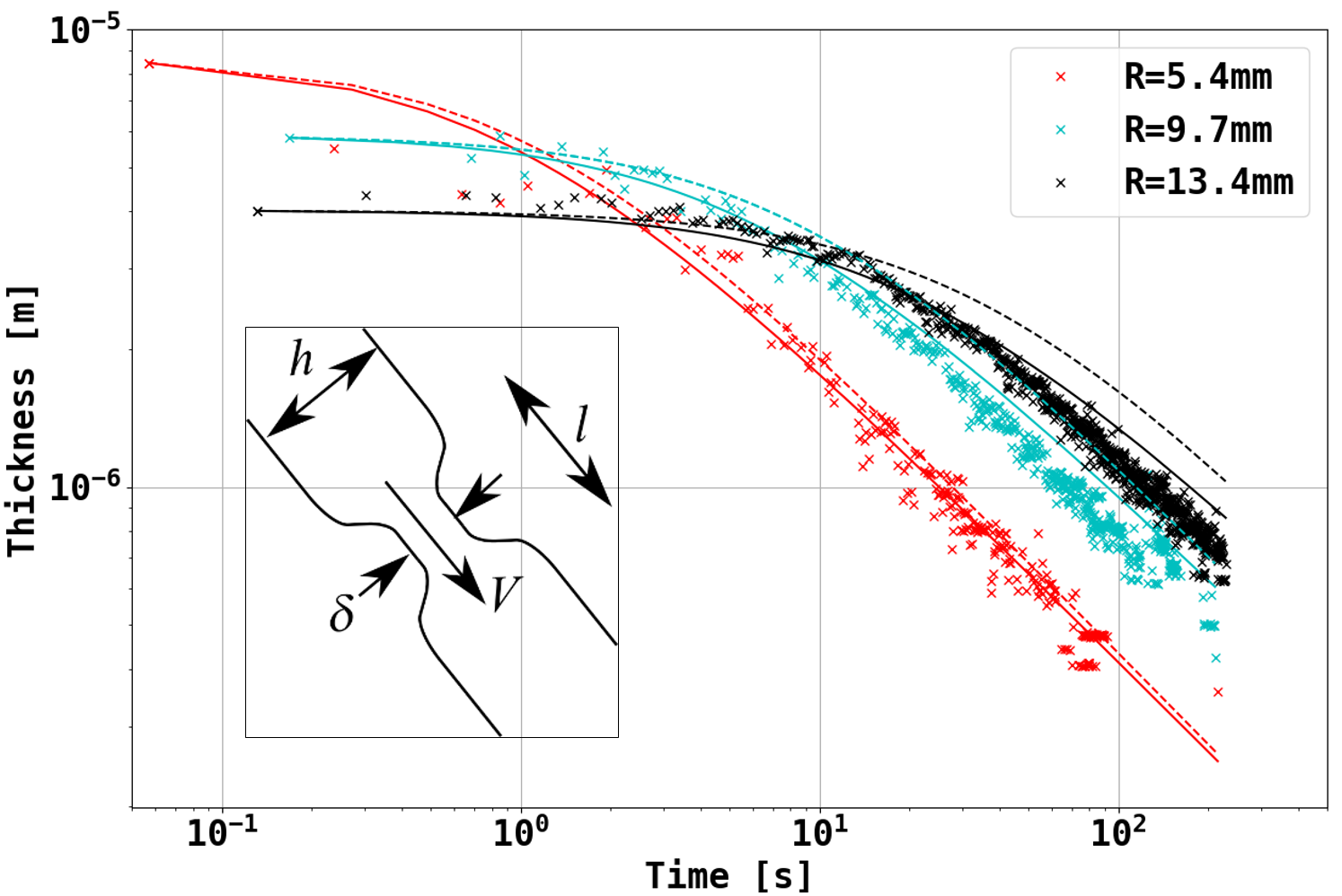}
      \caption{Thickness evolution curves of the film as a function of time for different bubbles size, close to water vapor saturation. The dotted lines represent the purely capillary drainage model, while the full lines additionally takes gravity into account.}
    \label{fig:Fig5}
\end{figure}
The system of equations for the drainage is therefore closed and leaves us with the possibility of a numerical integration of equation \ref{mass_conservation}, with $V$ given by equation \ref{velocity} and using $h_{0}$, the measured initial thickness, as an initial condition. Figure \ref{fig:Fig5} illustrates the validity of this approach with three representative bubbles (for the sake of
readability) taken from the same experiments as figure \ref{fig:Fig4}, when evaporation is negligible. Indeed the capillary model (dashed lines) underestimates the thinning velocity of the film as compared to the more complete model presented above that better describes our data. Of course the discrepancy between both models is more and more important as the bubble size increases. The remaining difference in terms of thinning rate
with the actual observations may be due to the non complete saturation of the water vapor, that would otherwise
lead to bubbles standing for $\sim 10^{4}-10^{5}$ s, as is developed further.

\subsection{Role of evaporation}
\label{subsec:evaporation}

\begin{figure*}[!ht]
  \centering
    \includegraphics[width=1\linewidth]{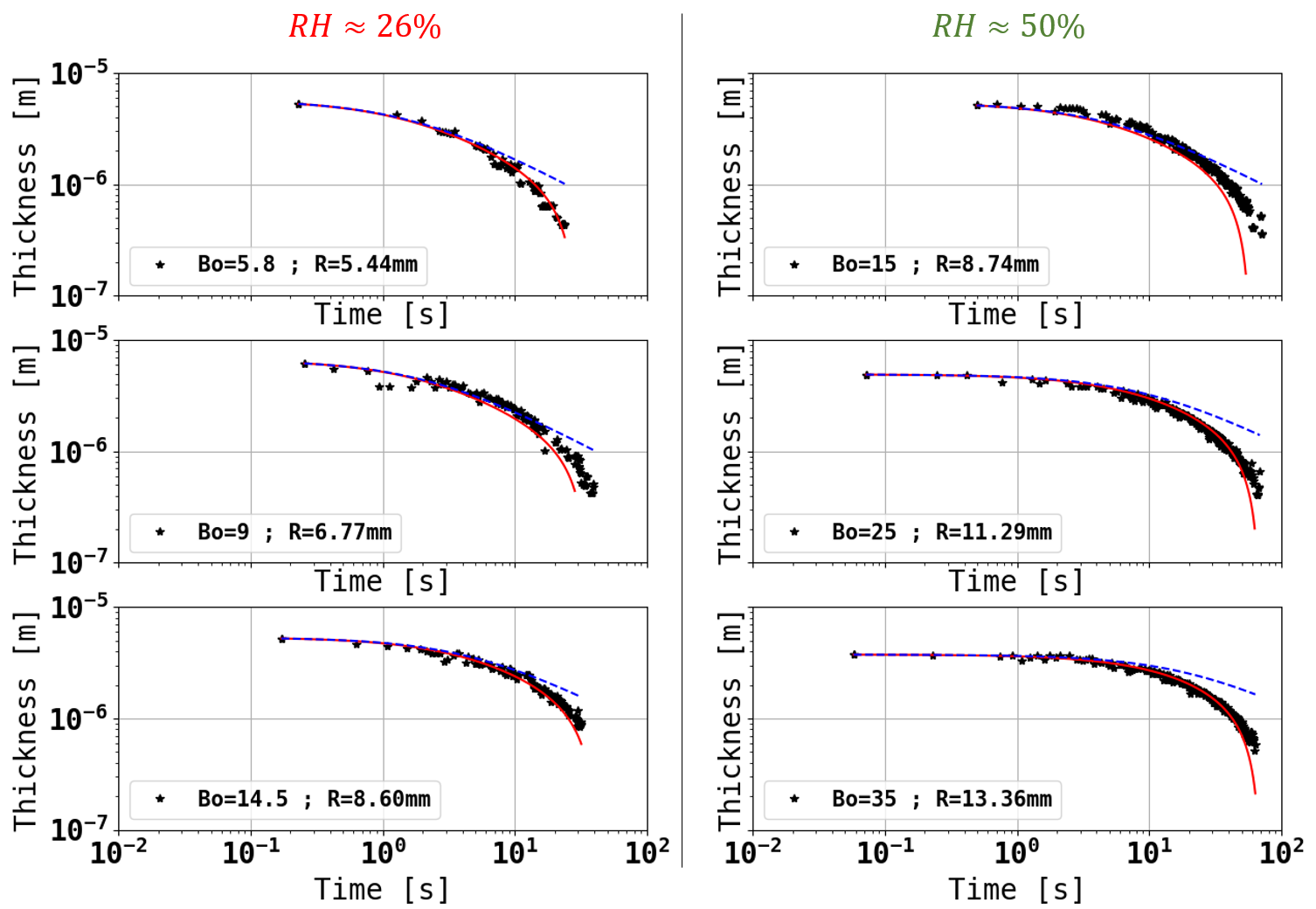}
  \caption{Thickness of bubbles cap as a function of time for various cap radii and relative humidity. The blue dotted lines represent the model without evaporation (section \ref{subsec:drainage}), and the red full line takes evaporation into account (section \ref{subsec:evaporation}).
  \label{fig:Fig6}}
\end{figure*}

With an appropriate model at hand to describe the thinning contribution due to the liquid flux from the cap to the bath, we now turn to the role of evaporation in this system. Our approach is inspired from Poulain \textit{et al.} \cite{Poulain2018} who first addressed this problem as such. Mass conservation must now account for this additional contribution and may be rewritten as follows:

\begin{equation}
    \rho_{\text{liq}} S\frac{dh}{dt}+\rho_{\text{liq}}PhV+SJ\sim 0,
	\label{mass_evap}
\end{equation}
where $J$[kg.m$^{-2}$.s$^{-1}$] is a mass evaporation rate that we need to estimate. To our knowledge, a complete model to describe the present situation that is, a 4 cm diameter bath over which a centimetric bubble is set has not been addressed yet and such a description is beyond the purpose of this paper. The limiting process is generally speaking the flux of water vapor in the gas phase, from the evaporating surface close to saturation, to infinity, where the relative humidity takes a fixed value (in our case, the setpoint of the PID controller). Water vapor being less dense than dry air, we must consider the possibility of a convection dominated evaporation. We calculate the Grashof number that balances the buoyancy of water-saturated air (that drives convection) and the viscous forces (diffusion) \cite{sanders1972franz}:
$Gr=|\frac{\rho_{\text{sat}}-\rho_{\infty}}{\rho_{\infty}}|\frac{gr_{\text{bath}}^{3}}{\nu_{\text{air}}^{2}},$
where $\rho_{\infty}$ and $\rho_{\text{sat}}$ are the density of air far from the bath and at saturation and are calculated from \cite{tsilingiris2008thermophysical}, $r_{\text{bath}}$ is the radius of our circular bath (2 cm) and $\nu_{\text{air}}\approx 1.5\text{x}10^{-5}$m$^{2}$.s$^{-1}$ the kinematic viscosity of air. Note that by using $r_{\text{bath}}$ as a characteristic lengthscale of the evaporating surface, we minimize the total evaporating surface (bath plus bubble) and therefore the convective effect. However, for a humidity of 50 \%, the Grashof is of 1528. We therefore consider that the boundary layer set by the evaporating bath is of primary importance. 
We make use of the scaling of Dollet \textit{et al} \cite{Dollet2017} for the convective evaporation of a circular bath:
\begin{equation}
    J_{\text{conv}}\approx \rho_{\text{air}}\frac{D}{r_{\text{bath}}}Gr^{1/5}\frac{M_{\text{liq}}}{M_\text{air}}\frac{P_{\text{sat}}}{P_{\text{0}}}(1-RH),
	\label{mass_evap_convection}
\end{equation}
where $D\approx 2\text{x}10^{-5}$ m$^{2}$.s$^{-1}$ is the diffusion coefficient of water vapor in air. The underlying approximation is that we neglect the influence of the bubble on the vapor concentration field (Figure \ref{fig:Diffusion-Convection}).(b). 
\begin{figure}[!ht]
  \centering
    \includegraphics[width=1\linewidth]{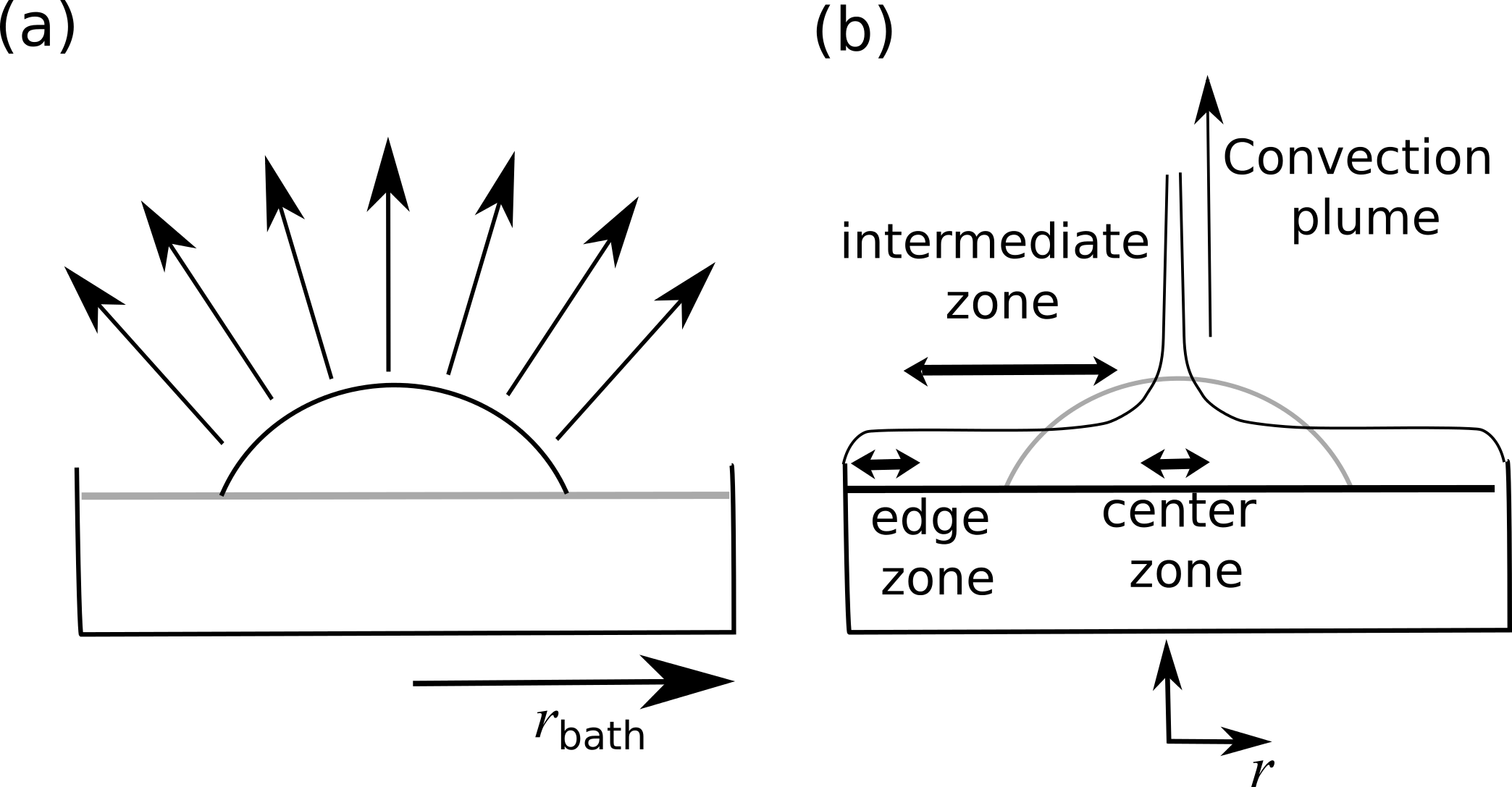}
  \caption{Scheme of the evaporation (a) in the diffusive limit, where only the evaporation of the bubble is considered and (b) in the conductive limit, where only the bath evaporation is considered.}
  \label{fig:Diffusion-Convection}
\end{figure}
In reference \cite{Dollet2017}, the calculation is performed in three different zones a central zone at $r<<r_\text{bath}$, an intermediate zone for $r<r_\text{bath}$ and a third zone for $r \approx r_\text{bath}$, $r$ being the radial coordinate with respect to the center of the bath.
We use the scaling of the intermediary zone, with $r<r_\text{bath}$.
Indeed, the spatial extension of the third zone is always smaller than 6 mm in our experiment. This explains why this zone is not relevant in this problem. 
Moreover, the area of the central zone, which radius scales as $Gr^{-3/5}r_\text{bath}$ represents 0.06 \% of the bubble cap area for $RH=50 \%$, which is negligible and justifies that we do not take into account the central zone either.

Figure \ref{fig:Fig6} shows the thickness evolution curves for bubbles of different radii and in varying humidity conditions. The model accounting for evaporation predicts remarkably well the dynamics of the system (solid red lines), with no adjustable parameters. We also show that evaporation plays a crucial role in the thinning of the bubbles at long times (the dotted blues lines represent the prediction of the model without evaporation), which eventually sets the overall bubbles stability.

It should be noted that the model does not recover the exponential law for the drainage, which was proposed by Champougny \textit{et al.} \cite{Champougny2016} for experiments obtained at 50\% humidity. The experiments are not in contradiction since the data presented in Figure 6 can also very well be fitted by an exponential. Our interpretation is that the apparent exponential obtained in \cite{Champougny2016} comes from from the addition of a power law and a constant evaporation rate.

We are aware that some models take into account evaporation inhomogeneities, which could generate stabilizing thermal Marangoni stresses \cite{Poulain2018a,dehaeck2014vapor}.
Nevertheless, the excellent agreement between the model and the data exhibited by the comparison in Figure \ref{fig:Fig6} shows that if such a stabilizing effect exists, it is a second order mechanism.

 Finally, the Grashof number depends on humidity. 
 In particular, in a almost saturated environment (RH = 99 \%), it is around 30. 
 For similar values of the Grashof number in evaporating drops, an approximate 50\% contribution of diffusion to the overall evaporation rate was reported\cite{Kelly2013}. 
 The presence of the bubble in the water vapor concentration field thus may become of importance at high humidity rate and we need to account for it. 
 In this case, we will thus make use of the model of the diffusive evaporation of a sphere (Figure 7.(a)), which appears to be in better agreement with our experimental data \cite{langmuir1918evaporation,Poulain2018a,fuchs1959evaporation}:
\begin{equation}
    J_{\text{diff}}=\rho_{\text{air}}\frac{D}{R}*\frac{M_{\text{liq}}}{M_{\text{air}}}\frac{P_{\text{sat}}}{P_{0}}(1-RH),
\label{evap_diffusion}
\end{equation}

\subsection{Prediction for the lifetime}
\label{subsec:lifetime}

\begin{figure}[!ht]
  \centering
    \includegraphics[width=1\linewidth]{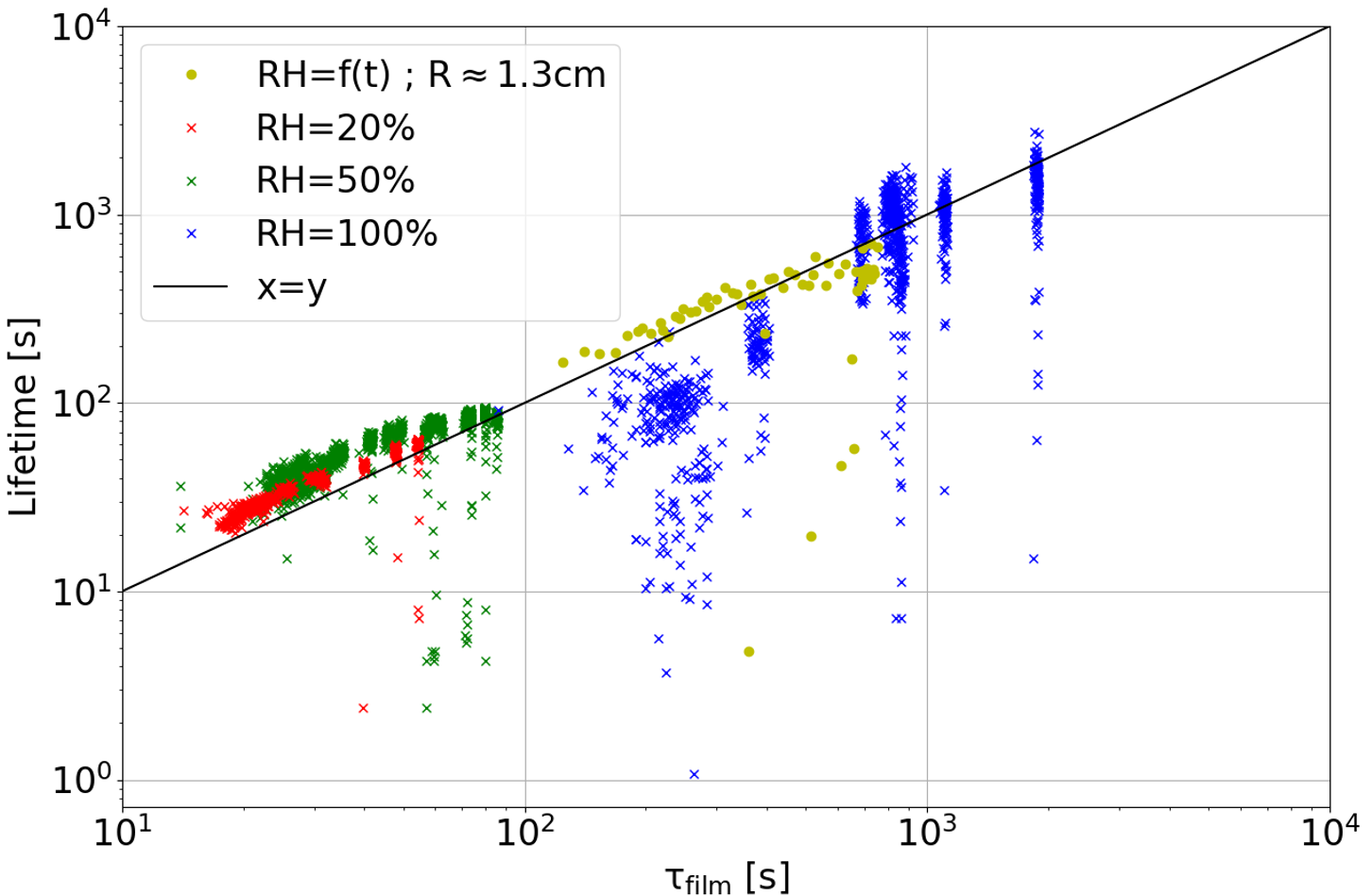}
  \caption{Measured lifetime of the bubbles as a function of the predicted lifetime presented in section \ref{subsec:lifetime}. Note that $\tau_\text{film}$ is calculated in presence of a diffusive evaporation for RH=100 \% and in presence of a convective evaporation for smaller RH, in line with the Grashof values calculated in both situations.}
  \label{fig:Fig7}
\end{figure}

In the following, the goal is to extract the bubbles lifetime $\tau_\text{film}$. The model presented above implies that the drainage of the bubble can be separated in two regimes. A first regime when the flow of liquid through the pinch dominates the global thinning of the cap film. And a second regime when evaporation thinning becomes dominant.

Combining equations \ref{mass_conservation} and \ref{velocity} and adding the evaporation flux, we get the complete mass conservation equation:
\begin{equation}
    \frac{dh}{dt}\approx -\frac{h}{R}\left(\frac{\gamma}{\eta}\frac{h^{3/2}}{R^{3/2}}+\frac{\rho_{\text{liq}}gh^{2}}{\eta}\right)-\frac{J}{\rho_{\text{liq}}},
	\label{mass_evap_bilan}
\end{equation}

Poulain \textit{et al.} \cite{Poulain2018a} proposed an analytical solution for the lifetime of the bubbles in the presence of capillary drainage only. In the case of the present situation of large bubbles, \textit{i.e.} $P/S \sim 1/R$, this lifetime can be expressed as:
\begin{equation*}
\tau_{\text{cap}}=R\left(\frac{\eta}{\gamma}\right)^{2/5}\left(\frac{\rho_\text{liq}}{J}\right)^{3/5} \times \int_0^{\infty} \frac{d\tilde{h}}{1+\tilde{h}^{5/2}} 
\end{equation*}
\begin{equation}
\sim 1.23 R\left(\frac{\eta}{\gamma}\right)^{2/5}\left(\frac{\rho_\text{liq}}{J}\right)^{3/5}.
\label{eq:LifeTimeCapillarity}    
\end{equation}

A similar calculation with only the gravity driven drainage leads to 
\begin{equation*}
\tau_{\text{grav}}=\left(\frac{\rho_\text{liq}R\eta}{g J^2}\right)^{1/3} \times \int_0^{\infty} \frac{d\tilde{h}}{1+\tilde{h}^{3}}
\end{equation*}
\begin{equation}
\sim 1.21 \left(\frac{\rho_\text{liq} R\eta}{g J^2}\right)^{1/3},
\label{eq:LifeTimeGravity}    
\end{equation}
\noindent a regime that is never reached since capillary driven drainage is always relevant.

In the following, we calculate the lifetime in presence of both a capillary and a gravity driven drainage.
To do so, we make use of an explicit Euler scheme to numerically integrate equation \ref{mass_evap_bilan} in order to retrieve the lifetime $\tau_\text{film}$ of the bubbles. The initial thickness of the film is taken constant and equal to 10 $\mu$m. It is higher than that shown in figures \ref{fig:Fig4}, \ref{fig:Fig5} and \ref{fig:Fig6} because the flow rate at which bubbles are inflated is higher in the lifetime measurement setup. Multiplying or dividing this value by a factor of 2 does not modify the results qualitatively. The rupture thickness is taken at 10 nm, consistent with a thermally induced instability of the film thickness that can lead to the final rupture \cite{Vrij1966}. Results are qualitatively unchanged if we take a rupture thickness of 100 nm.
The time necessary to achieve such thicknesses in the complete absence of evaporation results in an overestimation of the lifetime of the bubbles of at least two orders of magnitude. The reason for such discrepancy is that a complete saturation in water vapor is not achieved experimentally. We therefore arbitrarily chose an effective humidity value of 99\% for the "saturated" experiments (a slightly higher or lower value does not change qualitatively the results). For this saturated case, we thus use equation \ref{evap_diffusion} since the Grashof number is small, as explained in section \ref{subsec:evaporation} with RH = 99 \%.

Finally, as we argue in section \ref{subsec:evaporation}, we use the scaling for convective evaporation of the bath $J_\text{conv}$ (equation \ref{mass_evap_convection}) for all bubbles except that conducted at very high humidity for which the diffusive scaling \ref{evap_diffusion} is used.

Figure \ref{fig:Fig7} finally represents the measured lifetimes of the bubbles as a function of $\tau_\text{film}$. The model presented here thus provides a good representation of the bubbles lifetime both in terms of orders of magnitude and scaling behaviour with the bubbles size. The model predicts lifetimes slightly smaller than the observations. This is in qualitative agreement with the results on drainage shown in figure \ref{fig:Fig6}, where we can see that the prediction slightly overestimates the thinning rate. To emphasize the importance of convection, that predicts an evaporation flux independent of the bubbles size, we also represent in the Supplementary Information (see Appendix A) the same data inverting the use of equations \ref{mass_evap_convection} and \ref{evap_diffusion} to predict the evaporation. We immediately see that the scaling in terms of bubbles size for experiments far from the saturation is much worse which we think is the most convincing reason to consider the convective evaporation of the bath instead of the diffusive evaporation of the bubble. For the experiments close to saturation, the results suggest that the diffusive evaporation becomes important at this high humidity, since the scaling is better reproduced in this case (figure \ref{fig:Fig7}) than in the convective case (figure \ref{fig:Sup1}).

Finally, to address the importance of gravity on these systems, we made use of our numerical model to compare the predictions on the bubbles lifetime with ($\tau_\text{film}$) and without ($\tau_\text{cap}$) the second term in the right-hand side of equation \ref{velocity}, after making sure that the solution in the latter case indeed converges to the analytical solution given in equation \ref{eq:LifeTimeCapillarity}. We define a relative error as: 
\begin{equation}
 \text{Relative Error} = \frac{\tau_\text{cap}-\tau_\text{film}}{\tau_\text{film}}.   
\end{equation}
The results are plotted in supplementary materials (figure \ref{fig:Sup2}) and show that gravity accounts for up  to 15 \% of the predicted lifetime for the biggest bubbles, while it is completely negligible for the smallest ones.

\section{Conclusion}

In this work, we address the question of the stability of centimetric surface bubbles, made from a slightly concentrated surfactant solution. In particular, the impact of the size of the bubbles in this regime and the role of the environmental humidity are treated. We use an automated set up to repeat measurements in similar conditions in order to ensure the statistical significance of the results. The stability of the bubbles (their lifetime) was shown to increase both with increasing size and relative humidity. To explain these results, we measured the thickness evolution of the cap as a function of time. We derive a model for the thinning that accounts for both gravity and convective evaporation induced by the surrounding circular bath. Our procedure shows that a water vapor saturated environment is necessary to analyse the gravity/capillarity driven drainage for surfactant stabilised systems. With this model that successfully predicts the evolution of the cap thickness at hand, we perform an numerical integration of the evolution equation for the thickness of the cap and show that the lifetime thus predicted, assuming a burst thickness of 10 nm, is consistent with the experimental results. The fact that convective evaporation allows for a better prediction for the actual evaporation rate is an additional step towards predicting real systems where the surrounding pool of liquid is in general much larger than the bubbles (swimming pools, oceans, ...).

\section*{Acknowledgments} We acknowledge the technical contribution of Christophe Courrier for the electronic interface, David Hautemayou and Cédric Mézière for ensuring the mechanical support during the design of the experiment, Vincent Klein and Jérémie Sanchez for building the humidity controller. We are grateful to François Boulogne for fruitful discussions about evaporation and for the design of the oospectro library. We thank Cosima Stubenrauch and Nathalie Preisig for gracefully and very efficiently providing us with a detailed protocol for the recrystalization of TTAB. Funding from ESA (MAP Soft Matter Dynamics)
and CNES (through the GDR MFA) is acknowledged. This work was also funded by PepsiCo R\&D. The views expressed in this manuscript are those of the authors and do not necessarily reflect the position or policy of PepsiCo Inc.

\section*{Supplementary Information}

\subsection*{Comparison between the model with convection and evaporation}

Our main result is that a good description of the data necessitates to take into account an evaporation driven by the convection for RH = 20 and 50 \% and by diffusion in a saturated environment.

To make our point, we plotted in Figure \ref{fig:Sup1} the same data than in Figure \ref{fig:Fig7} using different values for $\tau_\text{film}$, which is calculated with equation \ref{evap_diffusion} for RH = 20 and 50 \% and with equation \ref{mass_evap_convection} for RH =100 \%. The scaling is much less convincing than in Figure 7, which shows that our description of evaporation catches better the main physical mechanisms.

\begin{figure}[!ht]
  \centering
  \includegraphics[width=1\linewidth]{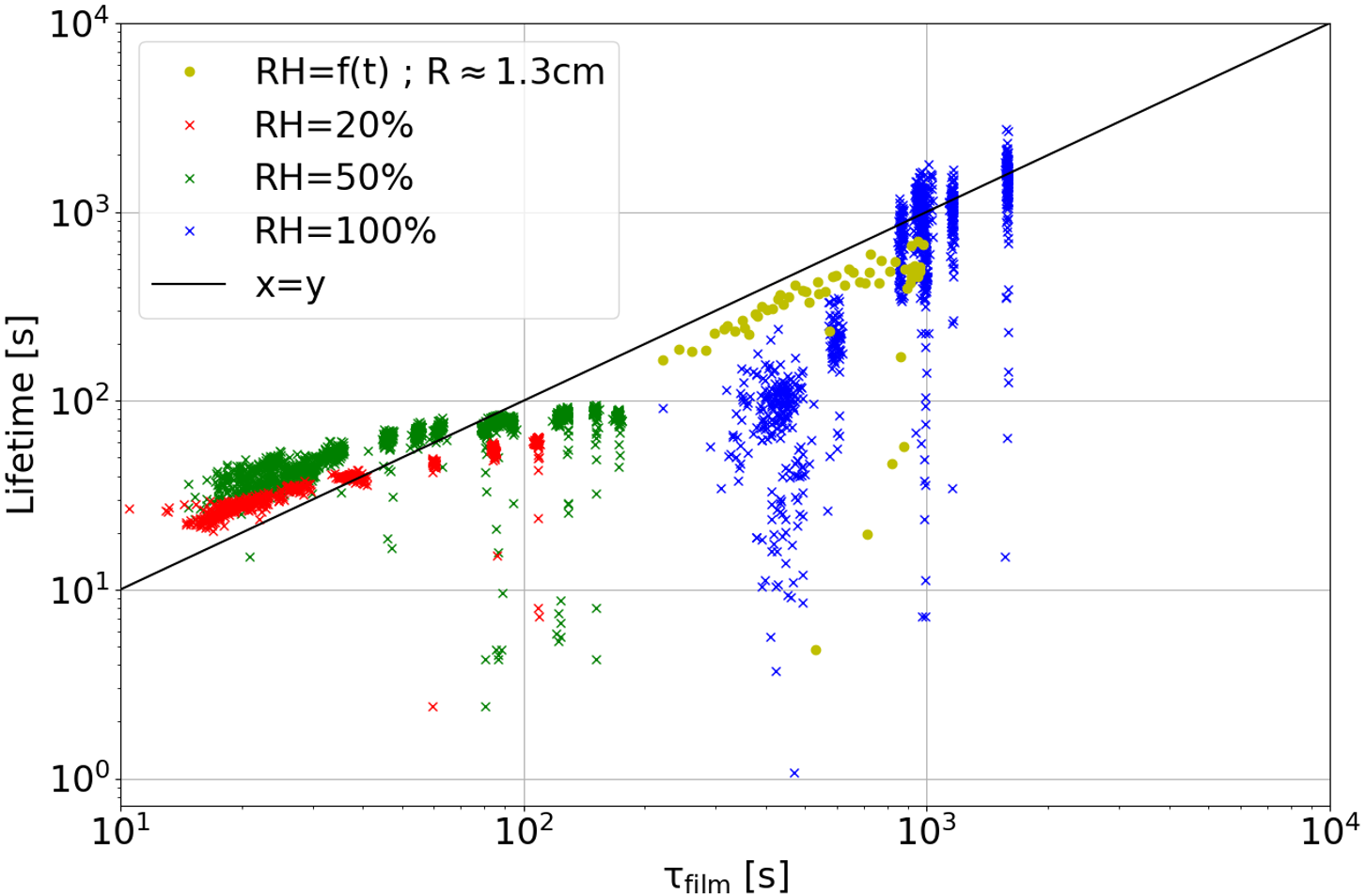}
  \caption{The data plotted in the figure are the same than in Figure 7. $\tau_\text{film}$ is now calculated in presence of a diffusive evaporation for smaller RH and in presence of a convective one for RH = 100 \%.}
  \label{fig:Sup1}
\end{figure}

\subsection*{Importance of the gravity driven drainage in the lifetime prediction}

We showed that the drainage curves are better described if the gravity drainage is taken into account. In Figure \ref{fig:Sup2}, we plotted the quantity $\frac{\tau_\text{cap}-\tau_\text{film}}{\tau_\text{film}}$, where $\tau_\text{cap}$ is given by Equation \ref{eq:LifeTimeCapillarity} and $\tau_\text{film}$ is calculated as explained in section \ref{subsec:lifetime}, versus the bubble radius for all our measurements. The result is that Equation \ref{eq:LifeTimeCapillarity} overestimates the lifetime by 5-10 \%. As expected, the overestimation grows with the bubble size. 

\begin{figure}[!ht]
  \centering
    \includegraphics[width=1\linewidth]{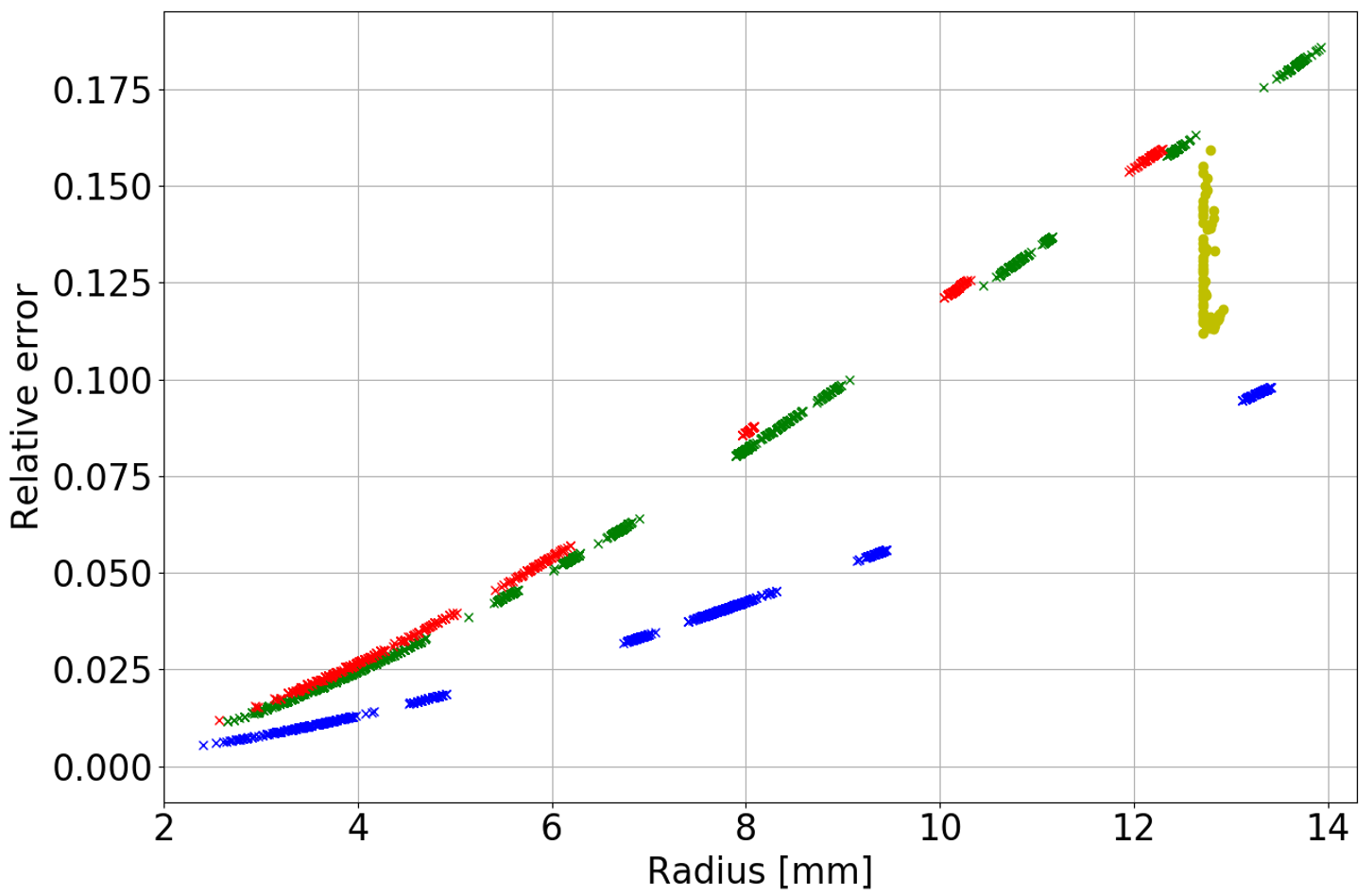}
  \caption{The relative error made when the lifetime is calculated with Equation \ref{eq:LifeTimeCapillarity}, \textit{i.e.} without gravity driven drainage, is evaluated by $\frac{\tau_\text{cap}-\tau_\text{film}}{\tau_\text{film}}$ and plotted versus the bubble radius for every experiment.} 
  \label{fig:Sup2}
\end{figure}

\bibliographystyle{unsrt}
\bibliography{Biblio}
\end{document}